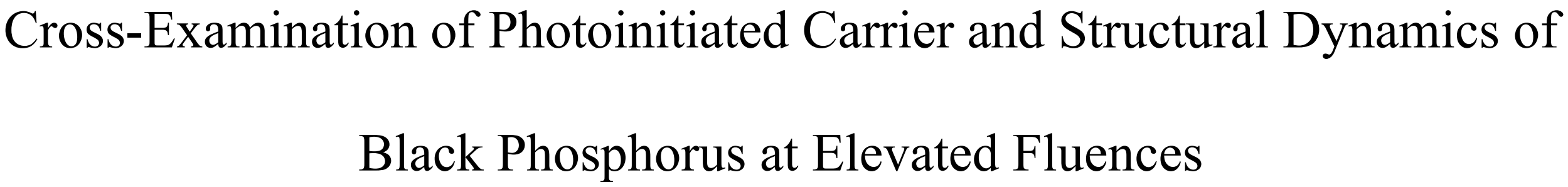

# Cross-Examination of Photoinitiated Carrier and Structural Dynamics of Black Phosphorus at Elevated Fluences

Mazhar Chebl,[†] Xing He,[†] Ding-Shyue Yang*

*Department of Chemistry, University of Houston, Houston, Texas 77204 United States*

[†]These authors contributed equally to this work.

*To whom correspondence should be addressed. Email: yang@uh.edu

**Abstract**

Revived attention in black phosphorus (bP) has been tremendous in the past decade. While many photoinitiated experiments have been conducted, a cross-examination of bP's photocarrier and structural dynamics is still lacking. In this report, we provide such analysis by examining time-resolved data acquired using optical transient reflectivity and reflection ultrafast electron diffraction, two complementary methods under the same experimental conditions. At elevated excitation fluences, we find that more than 90% of the photoinjected carriers are annihilated within the first picosecond (ps) and transfer their energy to phonons in a nonthermal, anisotropic fashion. Electronically, the remaining carrier density around the band edges induces a significant interaction that leads to an interlayer lattice contraction in a few ps but soon diminishes as a result of the continuing loss of carriers. Structurally, phonon–phonon scattering redistributes the energy in the lattice and results in the generation of out-of-plane coherent acoustic phonons and thermal lattice expansion. Their onset times at ~6 ps are found to be in good agreement. Later, a thermalized quasi-equilibrium state is reached following a period of about 40–50 ps. Hence, we propose a picture with five temporal regimes for bP's photodynamics.

## 1. Introduction

As a van der Waals (vdW) layered material, black phosphorus (bP) has intrigued scientists and engineers because of its various anisotropic behaviors along all three crystallographic axes [1]. Important steady-state properties were experimentally and theoretically investigated in detail in the 1980s by a few Japanese groups, where physical constants such as the effective masses of electrons and holes [2,3], the bandgap and carrier mobilities and activation energies [4,5], the dielectric constants [6], the pressure dependence of the bandgap [7], the specific heat and elastic constants [8,9], and the bulk modulus [10] were determined. Following the isolation of graphene and transition metal dichalcogenide as two-dimensional materials, the revived interests in bP have grown tremendously since 2014 [11-15], with exploration of the opportunities for electronic and optoelectronic applications and beyond [16-20]. In particular, photoinduced responses of bP have been examined using different techniques to shed light on the carrier dynamics with relatively low [21-24] to higher injection densities [25-27] as well as the resulting Burstein–Moss band-filling and bandgap renormalization effects [28-30]. While photocarriers in bP exhibit highly anisotropic charge transport especially in the in-plane armchair and zigzag directions [21,31], the carrier–carrier scattering and thermalization times appear to be largely insensitive to the different axes, which has been considered as a result of fast randomization of the carriers' distribution in *k*-space [22,23]. As to carrier–phonon coupling, time-resolved transient optical data contain oscillatory signals that indicate the generation of out-of-plane coherent acoustic phonons (CAPs) by a photoinduced strain impulse or breathing motions in ultrathin films [23,27,32]. A number of the aforementioned reports examined the layer dependence of the photoresponses, considering the strong influence of layer number and stacking order on bP's band structure [33-35] and their potential impacts on the dynamics.

However, even with these time-resolved optical studies, knowledge of bP's photoinduced structural dynamics only became available recently by using ultrafast electron microscopy, diffraction, and scattering methods [36-39]. In the in-plane directions, derived from the time-dependent changes of Bragg diffraction intensities, the initial rise in the atomic mean-square displacements (MSD) with a time constant of ~0.5 ps was assigned to electron–phonon equilibration related to the above-gap excess energy, whereas the slower component of ~20 ps with high anisotropy along the armchair and zigzag axes was attributed to phonon–phonon thermalization [37]. The following study of changes in the diffuse scattering suggested the relaxation and anisotropic distribution of photocarriers in bP's Brillouin zone in the first few ps, which results in the preferential emission of high-energy optical phonons with momenta along the zigzag direction [38]. Along the out-of-plane vdW-stacked direction, our previous report using ultrafast electron diffraction (UED) in reflection geometry revealed a unique carrier-coupled interlayer lattice contraction accompanied by symmetric $A_g$ coherent intralayer atomic vibrations in first few ps [39]. The strong dynamic correlation between the electronic band structure at the Z point (with the presence of carriers) and the layered lattice of bP was found to be of critical importance, which appears to be in line with the strong "quasi-covalent" interlayer coupling found at the steady state (without carriers) by Raman scattering and in calculation [40-42].

In this report, we further emphasize the importance of cross-examining bP's photoinitiated carrier and structural dynamics obtained at comparable experimental conditions. A consistent picture for bP's photoresponses can then be reached and resolve questions in the literature regarding, e.g., the fate and lifetime of photoinjected carriers, the nature of bP's anisotropic structural responses, and the temporal regimes for the dynamics. Theoretical models are also included in our analyses to quantify the experimental observations. We find that at elevated injection densities of $10^{20}$–$10^{21}$ cm$^{-3}$, most photocarriers are extremely short-lived and transfer their energy to the phonon dynamics almost immediately. The order(s)-of-magnitude-lower residual carrier density at the Z point plays the critical role of electronic–lattice dynamic coupling for the initial interlayer contraction. After several stages of nonthermal behavior and phonon thermalization, the lattice expansion observed by UED is found to contain both the stress-free temperature increase and stress-induced thermoelastic components, which gives a consistent connection to the CAP signals in the optical results.

## 2. Results

### 2.1. Energetics of photoinjected carriers in bP's band structure

A careful consideration of bP's electronic band structure and the experimental conditions is needed for a better understanding of bP's time-dependent photoresponses. In Figure 1, carrier injection made by an above-gap transition with 515-nm photons (2.41 eV) takes place over a broad range in the Brillouin zone. For semiconducting materials, the early-time ultrafast dynamics are often discussed in the context of thermalization of the energetic photocarriers toward the band edges and transfer of the above-gap excess energy via carrier–phonon coupling. The initial injection density at the surface is typically calculated by $N_0 = F_{\mathrm{ex}}(1 - R)\alpha/E_{\mathrm{ex}}$, where $F_{\mathrm{ex}}$ is the apparent fluence and $R$ and $\alpha$ are the reflectivity and the absorption coefficient at the excitation energy $E_{\mathrm{ex}}$, respectively; an exponential decrease inside the bulk is generally assumed to be the depth profile of the carrier distribution. However, $N_0$ may reach the level of $10^{20}$–$10^{21}$ $\mathrm{cm}^{-3}$ at a fluence of sub- to a few $\mathrm{mJ/cm^2}$ on bP. While bulk bP exhibits essentially a single-band behavior in a relatively large energy range of more than 1 eV across the bandgap $E_{\mathrm{g}}$ = 0.33 eV [4], the extent of carrier occupancy and accompanying effects due to band-filling at such an injection level can have major impacts on the observed photoresponses and carrier lifetime.

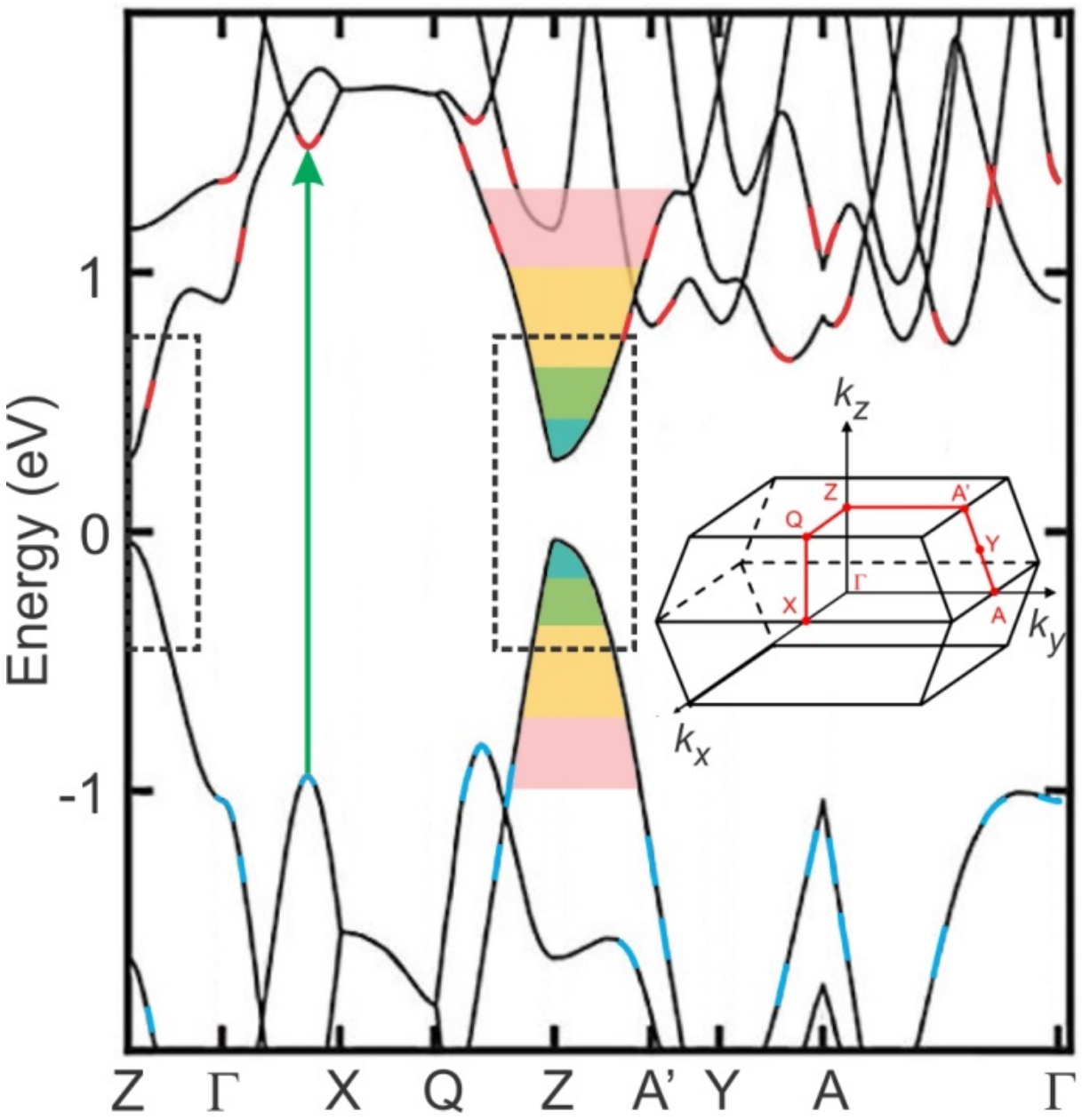

**Figure 1.** Electronic band structure of bulk bP (adapted from Ref. [38]) and relevant experimental conditions. Regions for the vertical transitions (green arrow) and the injections of electrons (red) and holes (blue) by 2.41-eV photons are denoted. The dashed boxes for an energy range of 1.2 eV indicate a quasi-single-band behavior centered at Z. The blue, green, yellow, and pink regions indicate the Fermi energy levels for band-filling of carrier densities of 0.3, 1, 3, and $5\times10^{20}$ $\mathrm{cm}^{-3}$, respectively, based on the parabolic band model. The inset shows the Brillouin zone with high symmetry points.

The density-of-state effective masses $m_{\mathrm{D}} = \sqrt[3]{m_a m_b m_c}$ (i.e., the geometric average of the effective masses along bP's three principal axes) of electrons and holes are $0.22m_0$ and $0.24m_0$, respectively, where $m_0$ is the mass of an electron [3]. Hence, the respective effective densities of states $2(2\pi m_{\mathrm{D}} k_{\mathrm{B}} T/h^2)^{3/2}$ near the conduction and valence band edges are $2.5\times10^{18}$ $\mathrm{cm}^{-3}$ and $2.8\times10^{18}$ cm$^{-3}$, where $h$ is the Planck constant, $k_{\mathrm{B}}$ the Boltzmann constant, and $T$ = 293 K is the base temperature. According to the Nilsson approximation for parabolic bands [43], the Fermi energy levels of electrons ($E_{\mathrm{Fc}}$) and holes ($E_{\mathrm{Fv}}$) are each ~1 eV away from their respective band edges for $N_0$ ~ $5\times10^{20}$ cm$^{-3}$ (calculated with $F_{\mathrm{ex}}$ = 3.23 mJ/cm$^2$), reaching the energy range beyond the single-band region (Figure 1). Multiple scenarios of dynamics are possible in such a highly degenerate condition. The extreme cases are ($i$) Photocarriers have a relatively long lifetime, and therefore a major portion of their above-gap excess energy is retained in the electronic subsystem, resulting in longer times for carrier relaxation due to intervalley scatterings and Pauli blocking and for the thermalization between carriers and the lattice; and ($ii$) photocarriers are short-lived due to, e.g. Auger and electron–hole recombination, exciton–exciton annihilation, and/or defect trapping, causing their energy to transfer to the lattice subsystem effectively that leads to a temperature increase following phonon thermalization. The actual situation could be somewhere in the middle.

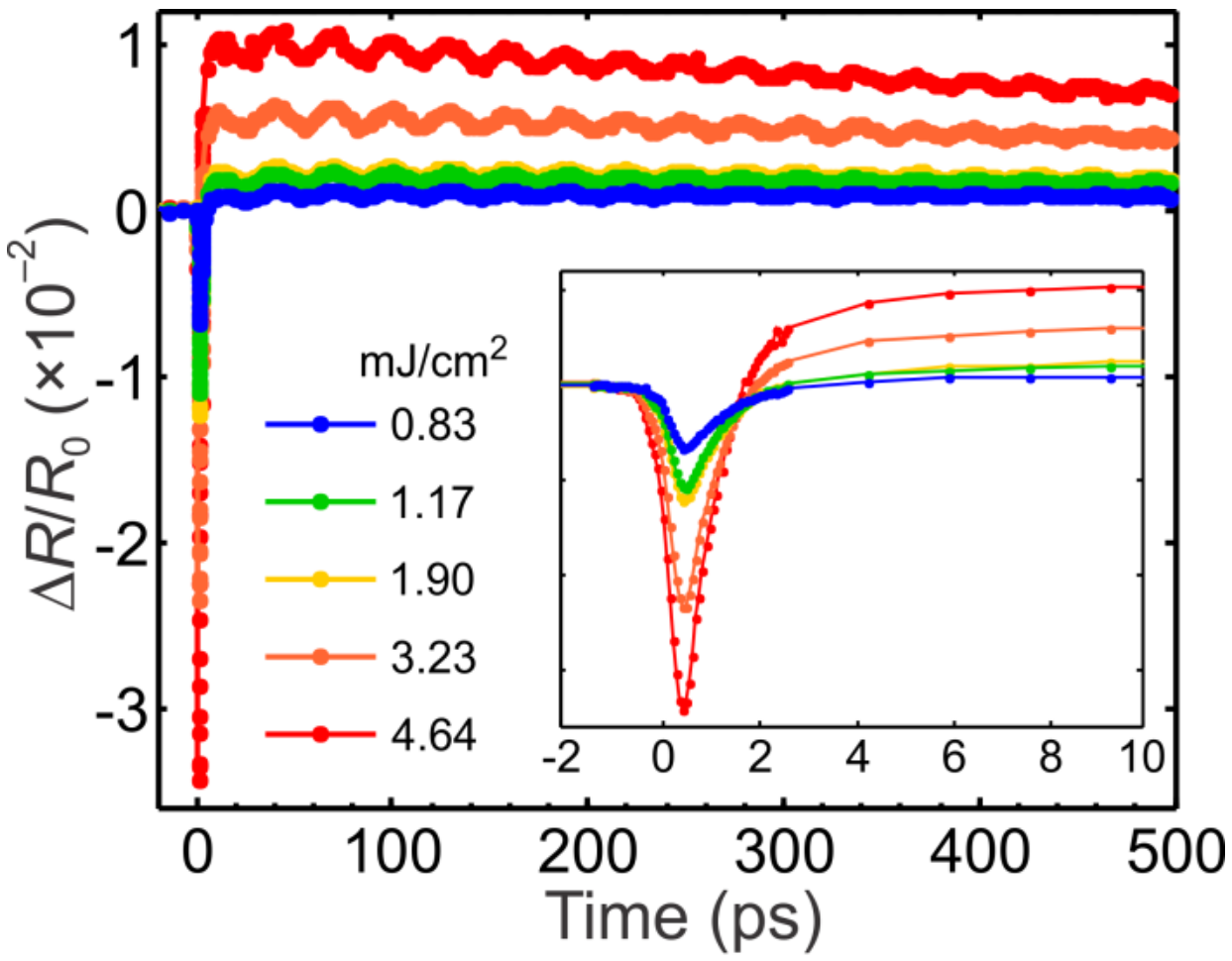

**Figure 2.** Optical transient reflectivity of bP acquired with the polarization of the 1030-nm probe pulses being along the zigzag $a$ axis. The fluence-dependent results show the dynamics with an initial drop and a fast rise followed by persisting oscillatory signals over a slowly diminishing background at longer times. The lines are guides to the eye. The inset shows the dynamics in the first 10 ps.

We argue that optical studies such as transient reflectivity (TR) or absorption measurements in the visible/near-infrared range may not always give a definitive answer especially when multiple time constants are involved. In the fluence range used here (0.8 to 4.6 mJ/cm$^2$), bP's initial TR, $\Delta R/R_0$, would be more than 20% decrease at 3.23 mJ/cm$^2$ and highly dominated by the contribution of free-carrier absorption, based on the theory that includes the Burstein–Moss band-filling and bandgap renormalization effects [44,45]. However, the experimental observation is apparently much more moderate, about an order of magnitude smaller for the initial decrease (Figure 2a). Such a discrepancy needs to be reconciled. A challenge is that both carrier annihilation processes and fast carrier diffusion can lead to a significant reduction of the surface carrier density. It is noted that bP has reasonably large carrier mobilities along all crystallographic directions including the cross-layer one [4], which may cause a decrease in the surface density by 50% within the first few ps if most photocarriers survive; the actual reduction could be even larger at a faster speed (up to 75% in 3 ps) due to the density-dependent $D$ value at the range of $10^{20}$–$10^{21}$ cm$^{-3}$ (see Figure S1 and further discussion in Supplementary Information) [46-48].

Nevertheless, as will be shown later especially with the UED results, the dominant carrier dynamics at initial times are in fact carrier recombination and energy transfer to phonons [37-39]. The first clue comes from the calculation of the carrier density at which the free-carrier absorption better matches with the experimental level, where about 90% of $N_0$ needs to be annihilated not very long after the instrumental response time (Figure 2, inset). Hence, the Z valley can sufficiently accommodate the greatly reduced carrier density following their fast thermalization without much involvement of other valleys (Figure 1). Furthermore, the majority of the remaining carriers will decay in a short period. At ~$10^{19}$ cm$^{-3}$, a carrier recombination time of 12–18 ps [22], 6–16 ps [23], or faster [28] has been reported. Knowing that photoluminescence does take place for multilayer bP [49,50] and longer-time dynamics at a low injection density were previously visualized [21,31], we believe that only carriers near the band edges remain for a relatively long time, which have consequences in the electronic–lattice coupling discovered by UED (see later) [39].

### 2.2. Transient reflectivity of bP from a few ps to longer times: Coherent acoustic phonons

In a self-consistent consideration, the surface temperature jump (within ~10% overestimation) *after phonon thermalization* may be reasonably estimated by $\Delta T = N_0 E_{\mathrm{ex}}/\rho C$, where $\rho$ = 2.708 g cm$^{-3}$ and $C$ = 0.696 J g$^{-1}$ K$^{-1}$ are the mass density and specific heat of bP, respectively [51]. As a result, an interlayer lattice expansion is anticipated but should only be detectable by time-resolved diffraction such as UED. Optically, the temperature jump causes a long-lasting increase in TR due to the thermooptic modulation of bP's refractive index, whose positive coefficient $\partial n/\partial T$ is about $2\times10^{-4}$ $K^{-1}$ (Figure 2). Furthermore, the thermoelastic strain initially developed at the sample surface leads to the generation of CAPs that travel into the bulk at the longitudinal speed of sound $v_{s,\parallel}$, which produce oscillatory signals in TR because of the interference between the reflected probe beams from the sample surface and the propagating strain pulse (Figures 2 and 3) [52,53]. Such phenomena in bulk bP were previously reported using transient absorption measurements [23,32]. Here, instead of the anisotropy and polarization dependence of the oscillations, we focus on the mechanism and implications of the underlying structural deformations, which will be compared with the structural dynamics derived from the UED results.

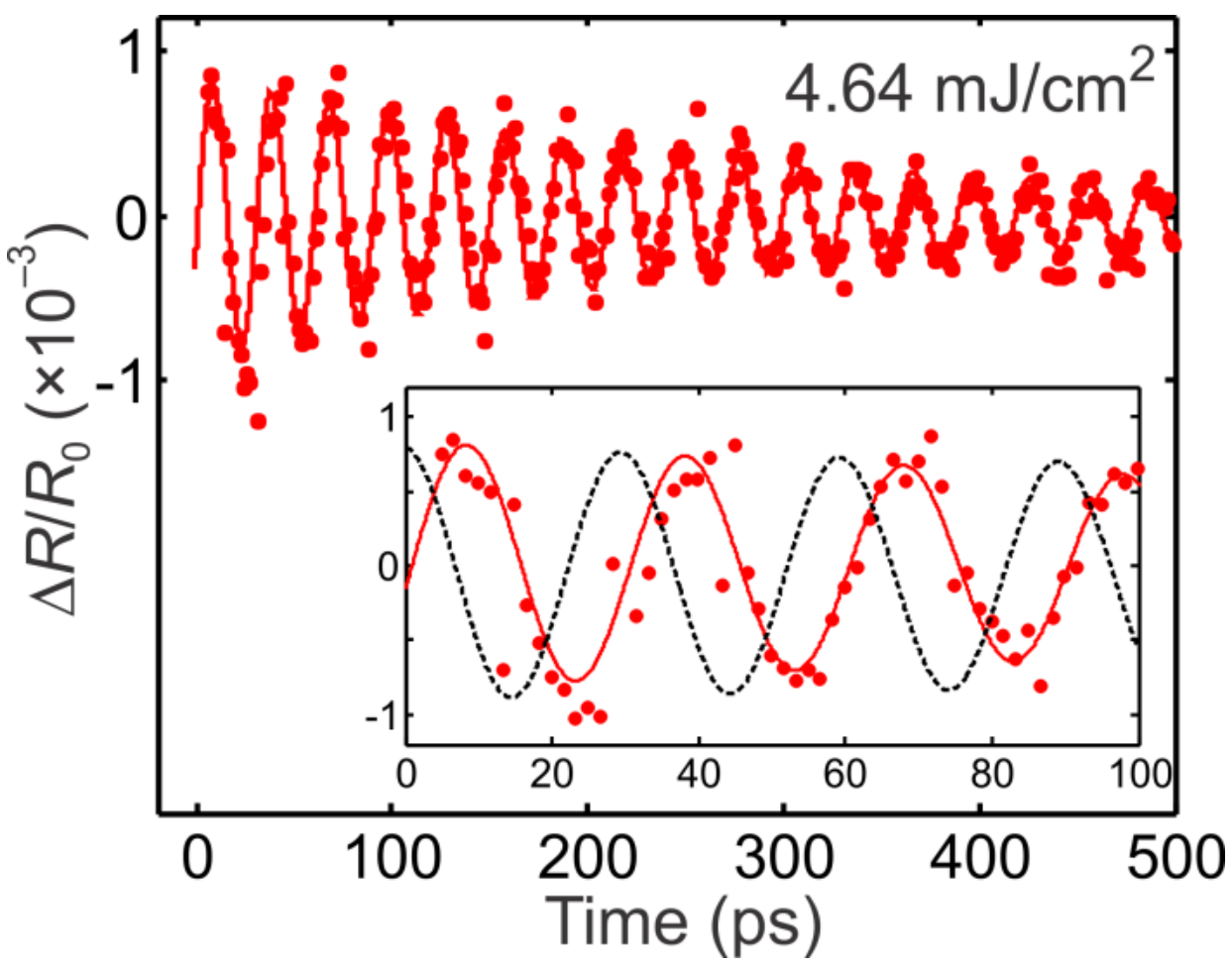

**Figure 3.** Oscillatory part of the transient reflectivity measured at 4.64 mJ/cm$^2$, which corresponds to the presence of long-lived coherent acoustic phonons propagating into the bulk. The red line is a fit based on Equation 1. The inset shows the first 100 ps, where the black dashed line is a theoretical curve assuming an instantaneous lattice expansion deformation. A temporal offset is noted.

In Figure 3, a fit of the periodic modulations to the model

$$\frac{\Delta R(t)}{R_0} = A_{\mathrm{osc}} \cos\left(\frac{2\pi t}{\tau_{\mathrm{osc}}} + \phi\right) \exp\left(-\frac{t}{\tau_d}\right) \tag{1}$$

yields the initial amplitude $A_{\mathrm{osc}}$ = $(8.31\pm0.08)\times10^{-4}$, the period $\tau_{\mathrm{osc}}$ = 29.85±0.07 ps, the phase $\phi \simeq 1.38$ (equivalent to an apparent delayed onset time of oscillation $t_{0,\mathrm{osc}} \simeq 6.6$ ps), and the dephasing time $\tau_{\mathrm{d}} \simeq 320$ ps. Almost the same results of $\tau_{\mathrm{osc}}$, $\phi$, and $\tau_{\mathrm{d}}$ are obtained at the other excitation fluences used. Hence, we obtain $v_{\mathrm{s},\parallel}$ = 5.36 km/s in good agreement with previous reports [9,23] according to the following equation [54,55],

$$\tau_{\rm osc} = \frac{\lambda_{\rm pr}}{2v_{s,\parallel}\tilde{n}_1} \tag{2}$$

where $\tilde{n}_1 = \sqrt{\epsilon_{\rm ZZ} - \sin^2\theta_{\rm in}}$ is the effective refractive index along the surface normal with $\theta_{\rm in}$ = 35° being the incidence angle of the probe beam and $\epsilon_{\rm ZZ}$ = 10.7+0.12$i$ being the dielectric constant along the zigzag axis at $\lambda_{\rm pr}$ = 1030 nm [6]. In addition, we find that the attenuation of the probe beam inside bP (with a decay length of $\lambda_{\rm pr}/4\pi\kappa_{\rm ZZ} \simeq$ 4.4 μm where $\kappa_{\rm ZZ}$ = 0.0186 is the imaginary part of $\tilde{n}_1$) can adequately account for the dephasing time, in that the difference between the paths of the reflected beams at time $\tau_{\rm d}$ is $2v_{\rm s,\parallel}\tau_{\rm d}/\cos\theta_{\rm in} \simeq$ 4.2 μm. This result, consistent with the previous report [23], signifies the induced CAPs to persist even on the ns scale or longer.

An analysis of $A_{\rm osc}$ and $\phi$ further reveals the optically detectable part of bP's photodynamics. First, we consider the model that assumes an instantaneous generation of the thermoelastic strain pulse (omitting the carrier-caused photoelastic contribution), whose TR amplitude is estimated by [56]

$$A_{\rm osc}(k) \cong \left| \frac{4ik}{1-\tilde{n}_1^2} a_{\rm cv} \frac{\partial \epsilon_{\rm ZZ}}{\partial E}\bigg|_{E=E_{\rm pr}} \frac{3B\beta\Delta T}{\rho v_{\rm s,\parallel}^2} \frac{2i\tilde{n}_1 k}{4\tilde{n}_1^2 k^2 + \alpha^2} \right|, \tag{3}$$

where $k \equiv 2\pi/\lambda_{\rm pr}$ is the probe beam wavenumber, $a_{\rm cv}$ the deformation potential including both the valence- and conduction-band contributions, $E_{\rm pr}$ the probe photon energy, $B$ the bulk modulus, and $\beta$ is the interlayer linear thermal expansion coefficient *in the absence of stress*. We obtain a theoretical value of $A_{\rm osc} \simeq 8.46\times10^{-4}$, which agrees satisfactorily with the experimental result (see Table S1 for the physical constants used). Additionally, the longitudinal thermoelastic strain at the surface *due to thermal stress* is given by $\eta_{\rm th} = 3B\beta\Delta T/\rho v_{s,\parallel}^2$ [56]. Regarding the phase of the oscillatory signals, the contributions come from (*i*) the real ($n_{\rm ZZ}$) and imaginary ($\kappa_{\rm ZZ}$) parts of $\tilde{n}_1$, $\phi_{nk} = \tan^{-1}\left[\kappa_{\rm ZZ}\left(n_{\rm ZZ}^2 + \kappa_{\rm ZZ}^2 + 1\right)/n_{\rm ZZ}\left(n_{\rm ZZ}^2 + \kappa_{\rm ZZ}^2 - 1\right)\right]$ (about $7.0\times10^{-3}$ in the current study), and (*ii*) the derivatives with respect to the strain, $\phi_{\rm strain} = \tan^{-1}[(\partial\kappa_{\rm ZZ}/\partial\eta)/(\partial n_{\rm ZZ}/\partial\eta)]$ (about 0.10) [52,54]. However, the sum is much lower than the experimental observation; the calculated TR plotted in the inset of Figure 3 also shows the clear temporal offset (See Supplementary Information for the equation). The origin of this onset difference may be explained by the unique structural dynamics of bP found by UED at initial times.

## 2.3. Cross-examination of structural and carrier dynamics of bP

As summarized in the Introduction about the anisotropic nonthermal structural dynamics of bP, we believe that details may require adjustments in light of the critical information obtained from the TR results about ultrafast carrier annihilation at the injection density of $10^{20}$–$10^{21}$ cm$^{-3}$ and a surviving density of $10^{19}$ cm$^{-3}$ or less within 1 ps. Although still due to electron–phonon coupling, the initial in-plane MSD increase in ~0.5 ps [37] is better characterized as the result of a nearly complete transfer of energy from the annihilation of photoinjected carriers to the lattice (even though $E_{\mathrm{g}}$ is a relatively minor fraction of $E_{\mathrm{ex}}$). The phonon–phonon coupling and thermalization are indeed highly anisotropic; however, since most photocarriers are extremely short-lived, the significance of their influence in the Brillouin zone for the anisotropic optical phonon population (Figures 2b of Ref. [38] in a few to few tens of ps) is called into question.

Along the vdW-stacked direction, the lack of apparent fluence dependence for the initial contraction of $8\times10^{-4}$ was intriguing [39], which now becomes clear when the initial order(s)-of-magnitude reduction of the carrier density and the short carrier lifetime are taken into account. This validates our previous view that the Lifshitz or superlattice model with a Drude-like electron gas in charged slabs [57] does not adequately explain the experimental observation because of the requirement for long-lived carriers at a large density of $>10^{20}$ cm$^{-3}$. Instead, a low density at the Z point is enough to cause a strong dynamic correlation with the layered lattice of bP. Furthermore, from the viewpoint of the involved $k$-space vector, we argue that the interlayer electronic–lattice coupling also gives rise to the emission of symmetric $A_g$ optical phonons, considering the interplay between the out-of-plane P–P bonds in the puckered layers and the electron–hole wavefunctions at the Z point [39]. We also note that direct coupling to acoustic phonons via the acoustic deformation potential is not prominent initially, as the theoretical lattice strain at the surface due to the electronic stress, $\eta_{\mathrm{el}} = -Na_{\mathrm{cv}}/\rho v_{s,\parallel}^2$ [53] is only $-1.2\times10^{-4}$ with $N \simeq 10^{19}$ cm$^{-3}$. However, the same sign of the surface strain, i.e. a contraction instead of an expansion, is predicted from the electronic–acoustic phonon coupling.

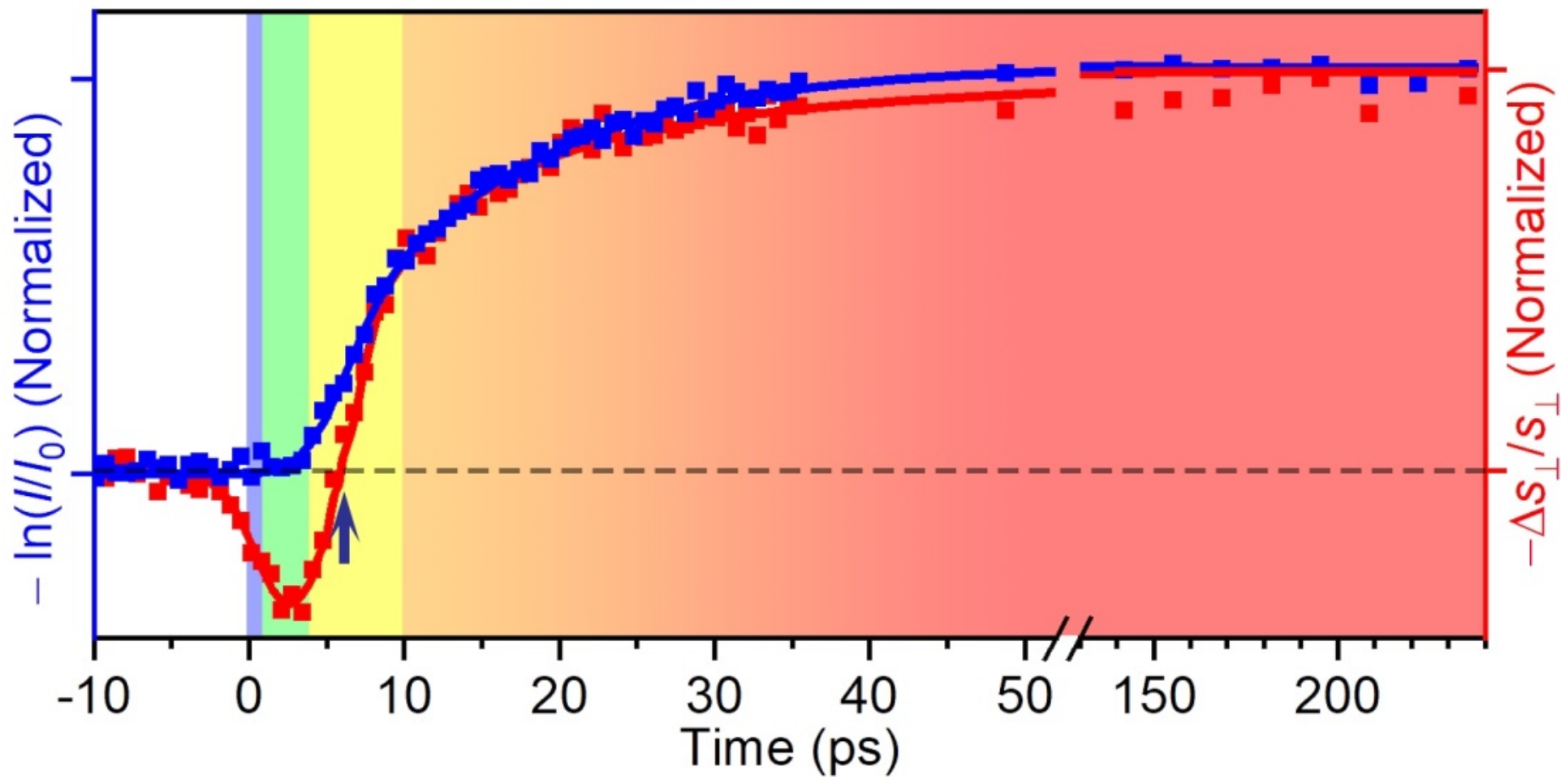

**Figure 4.** Photoinitiated changes of the (0 16 0) Bragg diffraction of bP. Different colors (light blue for the first ps, light green for 1–4 ps, yellow for 4–10 ps, orange from 10 ps transitioning to light red at about 50 ps) are used to indicate the five temporal regimes for bP's photodynamics (see text). The arrow indicates the onset time of lattice expansion following the reversal of lattice contraction.

Thus, up to five temporal regimes may be considered for bP's photodynamics (Figure 4). In the first regime of <1 ps, the ultrafast carrier annihilation (likely due to Auger recombination) and energy transfer via electron–phonon coupling are reminiscent of the behavior of a photoexcited elemental metal. However, the anisotropy of bP's bonding dictates the initial anisotropic population of optical phonons especially the in-plane ones [38], and therefore a phenomenological use of the two- or three-temperature model is not adequate. In the second regime up to 4 ps, the out-of-plane dynamics including an interlayer lattice contraction and coherent intralayer $A_g$ vibrations become prominent following the thermalization of the remaining carriers at the Z point. Afterward, in the third regime between 4 and 10 ps, further reduction of the carrier density cannot support the existing lattice contraction, while phonon–phonon scattering becomes more significant for later thermalization. Although still in a nonthermal state, it is during this time that a tensile thermoelastic strain near the surface and CAPs are developed, giving rise to the oscillatory TR signals with a delayed onset time of ~6.6 ps. Incidentally, the reversal time of the contraction to a lattice expansion also takes place at ~6 ps and does not change much by the fluence used (Figure 4, arrow). In the fourth regime to about 40–50 ps, bP's dynamics are mainly about the thermalization of phonons in all crystallographic directions, where Debye–Waller-type random atomic motions become mature, the diffraction intensities decrease to their minimum values, and the out-of-plane thermal expansion rises [37,39]. Finally, a quasi-thermal-equilibrium state is reached although a small density of photocarriers remains.

The amount of out-of-plane thermal expansion, e.g. $\Delta b \simeq 2.1$ pm at 2.4 mJ/cm$^2$, may also be better explained. It is noted that the corresponding surface temperature increase, $\Delta T$ = 76 K, in a stress-free condition cannot fully account for the observation: $\beta \Delta T \cdot b$ = 0.94 pm. For bP with long-lasting CAPs, the presence of the surface thermoelastic strain due to stress must also be included, which gives an additional contribution of $\eta_{\mathrm{th}} \cdot b$ = 1.3 pm. Hence, the theoretical prediction that includes both contributions agrees well with the experimental result. This finding further supports the strength of cross-examining time-resolved optical and structural data to obtain an informed and consistent picture.

## 3. Conclusion

We have shown that photoinitiated dynamics of bP and the relevant five time ranges during the thermalization process can be better understood by cross-examining its time-resolved TR and UED results, given the complementary nature of the two methods. TR with an optical probe informs about changes in a material's dielectric constant, induced mostly by carrier-related phenomena and also by the resulting lattice strain. In contrast, UED as a direct structure-probing method reveals picometer-level changes in the atomic positions and motions with an appropriate temporal resolution. In this report, it was shown crucial to use both sets of data to reach a self-consistent picture for bP's photocarrier and structural dynamics. We have found that at elevated laser fluences with injection densities at the level of $10^{20}$–$10^{21}$ cm$^{-3}$, photocarriers are found extremely short-lived and transfer all their energy to preferential phonon modes in the first stage. The remaining carriers near the direct-gap band edges exert significant electronic–lattice coupling that leads to an interlayer contraction with coherent intralayer vibrations in the second stage for a few ps. The continuing recombination of carriers and phonon–phonon scattering naturally result in the beginning of a thermal lattice expansion and the generation of coherent acoustic phonons in the third stage up to ~10 ps. The structure-probing UED further informs that atomic motions undergo thermalization for tens of ps, before a quasi-equilibrium state is reached in the final stage and the assignment of a temperature becomes valid. Given the updated knowledge about the carrier lifetime and the origin of structural changes, it is now important to examine how phonons couple anisotropically and thermalize to give bP's unique behavior.

## 4. Methods

Details about the TR and reflection UED apparatus [39,58] and the bP measurements [39] have been described previously. The ultrahigh vacuum (UHV) chamber assembly with a base pressure of $2\times10^{-10}$ torr allowed the preservation of an exfoliated bP surface's condition over an extended period, confirmed by no loss of diffraction intensities and optical reflectivity after multiple rounds of experiments. The UED instrumental response time was improved to ~500 fs with the implementation of a pulse-front tilt setup to overcome the temporal mismatch between the arrivals of the optical pump and electron probe beams. The TR signals from the same sample surfaces were initially recorded with this pulse-front tilt setup in the same UHV chamber. However, to achieve a better temporal resolution, additional data were acquired in a separate high vacuum chamber without the pulse-front tilt. These TR results were compared and found to follow the same dynamic changes.

## Acknowledgments

This research was primarily supported by the R. A. Welch Foundation (E-1860). The instrumental implementation of the pulse-front tilt scheme was partly supported by a National Science Foundation CAREER Award (CHE-1653903) and X.H. is partly supported by a National Science Foundation grant (CHE- 2154363).

## Conflict of interest

The University of Houston System requires the following statement: D.-S.Y. is a co-patent holder on US Patent No. 8,841,613. For the current TR and UED work, the authors declare no competing financial interest.

## ORCID

Xing He https://orcid.org/0000-0001-5341-5662

Ding-Shyue Yang https://orcid.org/0000-0003-2713-9128